\documentclass[sigconf, authorversion]{acmart}

\usepackage{hyperref}

\AtBeginDocument{%
  \providecommand\BibTeX{{%
    \normalfont B\kern-0.5em{\scshape i\kern-0.25em b}\kern-0.8em\TeX}}}

\copyrightyear{2024}
\acmYear{2024}
\setcopyright{rightsretained}
\acmDOI{10.1145/3649902.3655661}
\acmISBN{979-8-4007-0607-3/24/06}

\acmConference[ETRA '24]{2024 Symposium on Eye Tracking Research and
Applications}{June 4--7, 2024}{Glasgow, United Kingdom}
\acmBooktitle{2024 Symposium on Eye Tracking Research and Applications
(ETRA '24), June 4--7, 2024, Glasgow, United
Kingdom}

\begin{document}

\title{SARA: Smart AI Reading Assistant for Reading Comprehension}


\author{Enkeleda Thaqi}
\email{enkeleda.thaqi@tum.de}
\affiliation{%
  \institution{Technical University of Munich}
  \city{Munich}
  \country{Germany}
  \postcode{80333}
}

\author{Mohamed Mantawy}
\email{mohamed.mantawy@tum.de}
\affiliation{
  \institution{Technical University of Munich}
  \city{Munich}
  \country{Germany}
  \postcode{80333}
}

\author{Enkelejda Kasneci}
\email{enkelejda.kasneci@tum.de}
\affiliation{%
  \institution{Technical University of Munich}
  \city{Munich}
  \country{Germany}
  \postcode{80333}
}

\renewcommand{\shortauthors}{Thaqi, et al.}

\begin{abstract}
SARA integrates Eye Tracking and state-of-the-art large language models in a mixed reality framework to enhance the reading experience by providing personalized assistance in real-time. By tracking eye movements, SARA identifies the text segments that attract the user's attention the most and potentially indicate uncertain areas and comprehension issues. The process involves these key steps: text detection and extraction, gaze tracking and alignment, and assessment of detected reading difficulty. The results are customized solutions presented directly within the user's field of view as virtual overlays on identified difficult text areas. This support enables users to overcome challenges like unfamiliar vocabulary and complex sentences by offering additional context, rephrased solutions, and multilingual help. SARA's innovative approach demonstrates it has the potential to transform the reading experience and improve reading proficiency.
\end{abstract}



\begin{teaserfigure}
  \includegraphics[width=\textwidth]{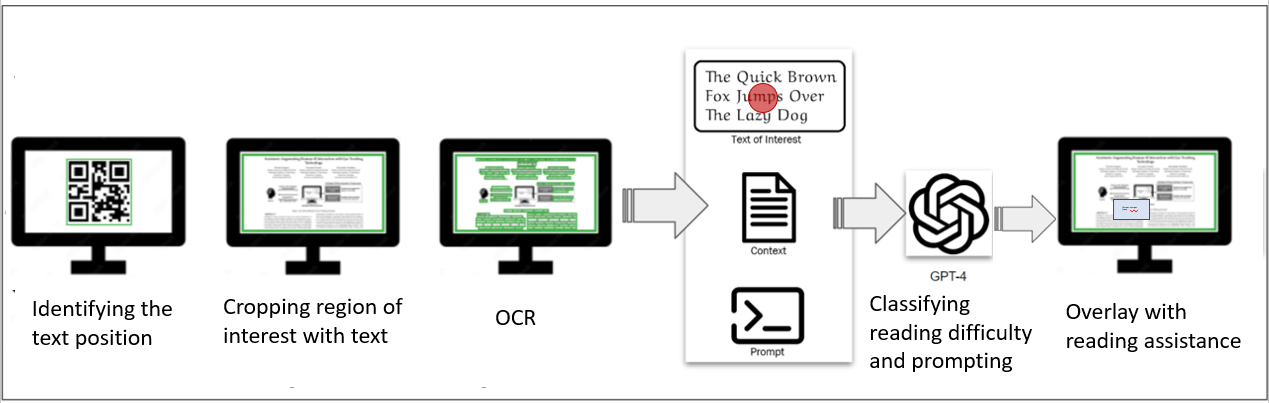}
  \caption{Illustration of key components SARA uses: text position identification, OCR, reading difficulty classification, prompt initiation for necessary help, and displaying assistance within the user's field of view. The workflow of the components is exemplified using the paper by \cite{langner2023teaser} as reading material.}
  \Description{SARA flow chart}
  \label{fig:teaser}
\end{teaserfigure}

\maketitle
\label{intro}
\section{Introduction}
Technological innovation is continuously redefining our interaction with our environment. Mixed reality (MR) stands out within these advances as a particularly promising framework in which physical and virtual dimensions blend seamlessly to create deeply immersive experiences \cite{gardony2020mr}. Within this context, the development of the Smart AI Reading Assistant (SARA) represents a significant breakthrough that aims to improve the reading experience.
SARA addresses common reading challenges by leveraging state-of-the-art head-mounted display (HMD) devices and advanced large language models (LLMs). 
This work explores the SARA application, demonstrating its technological framework and highlighting its ability to detect reading problems, extract text from the user's field of view, and provide personalized real-time support tailored to individual needs. Through a comprehensive exploration of SARA's functionalities, this research aims to show the potential of eye tracking, LLMs, and MR technology in improving reading experiences and increasing educational efficiency.

\label{background}
\section{Related Work}
\subsection{Eye Tracking in Reading Studies.} Eye-tracking technology is used in research studies to explore reading processes. It is commonly applied to investigate predictive abilities in reading comprehension \cite{meziere2023readcomp}, the cognitive load of reading and processing of read text \cite{delgado2022cognitveload}, and for the diagnosis of reading difficulties such as dyslexia \cite{nielson2016dyslexia}. Additionally, eye tracking serves as a suitable technology for the treatment and improvement of reading skills \cite{caldani2020visual}.
\subsection{LLMs and AI-based tools in Education.} \cite{kasneci2023chatgpt} explores the integration of LLMs into educational settings and highlights the potential advantages and challenges. The paper emphasizes the significance of critical thinking skills and advocates for clear strategies and pedagogical approaches to fully exploit the educational potential of LLMs while managing their complexity and limitations. \cite{vanwyk2024chatgpt} concludes that using AI-based tools presents an opportunity for innovative pedagogical approaches for students.
\subsection{AR/MR in Education.} In medicine, eye tracking and MR frameworks are used to train medical students and professionals on surgical skills \cite{lu2020armed}. In science, developing such frameworks enables the evaluation of novice and expert problem-solving skills during experiments and training novices based on the results \cite{sonntag2021science}.

\label{methods}
\section{Methods}
SARA was implemented and deployed for HoloLens2 \cite{hololens}. This MR-HMD device was chosen for its embedded features, such as integrated eye tracking, a built-in scene camera, and publicly available toolkits for Unity. These features provide a robust set of tools and resources that make it possible to provide a comprehensive solution for reading assistance in the MR environment.\\
The application operates through the following steps that are illustrated in Figure \ref{fig:teaser}:
\begin{enumerate}
    \item \textbf{Text position identification.} A QR code is displayed, and when detected and decoded, augmented virtual markers are placed in their respective positions in the real world. This approach guarantees the accurate identification of the text position containing the reading material the user is currently engaged with.
    \item \textbf{Text extraction.} In a multi-step approach, frames are taken by the HoloLens2 scene camera to capture visual data from the user's field of view for text processing. The focus is then narrowed down to the region of interest surrounding the virtual screen placed based on the QR code. The captured frames are then cropped to isolate the area containing the text being read.
    \item \textbf{Optical Character Recognition (OCR).} OCR is used to analyze the cropped images and extract text content. This process not only identifies text in the captured images but also generates bounding boxes around each recognized text element, providing precise information on the location of the text.
    \item \textbf{Gaze tracking for focus detection.} The user's gaze is captured in the MR environment, providing ray information indicating the location of the user’s focus on the text region, including the direction of gaze and subsequent eye movements.
    \item \textbf{Gaze alignment.} In this step, a series of coordinate transformations are performed. First, the point coordinates at which the user's gaze intersects with the screen are logged in relation to the world coordinates. Then, the coordinates of this point relative to the center of the virtual screen are calculated by subtracting the position of the virtual screen from the gaze intersection point. This step helps to create a local coordinate system centered around the virtual screen. Next, the intersection point, which is now in the coordinate system of the virtual screen, is transformed into the coordinate system of the cropped image by scaling the point based on the ratio between the width and height of the virtual screen in the world coordinate system and the width and height of the cropped camera image in the pixel coordinate system. This scaling ensures that the transformed coordinates accurately represent the position of the user's gaze on the region of interest in pixel units. Finally, the transformed coordinates are further adjusted to align with the top-left axis of the screen. To do this, the x-coordinate is shifted by half the width of the virtual screen in pixels, and the y-coordinate is mirrored to match the pixel orientation of the cropped image. The resulting transformed coordinates represent the exact pixel position of the user's gaze on the screen and allow the exact mapping between the user’s eye gaze and the position of the words in the text material.
    \item \textbf{Classifying reading difficulty.} SARA addresses two main challenges in reading: encountering unfamiliar words and difficulties comprehending paragraphs. Based on gaze dwell time, SARA identifies instances where users spend more time on unfamiliar words compared to the surrounding text, indicating potential difficulties. In addition, SARA detects regressions in reading patterns where users exhibit eye movements that do not follow the standard left-to-right and top-to-bottom progression. By using these detection techniques, SARA can identify areas of reading difficulty and offer immediate assistance.
    \item \textbf{Reading Support.} SARA uses the advanced features of state-of-the-art GPT-4 to overcome user's reading difficulties. For an unfamiliar word, SARA prompts a definition in context or a translation according to the user's needs to ensure accurate comprehension. Offering contextual definitions or translations improves comprehension and enables a smoother reading experience. When users have difficulty understanding paragraphs, SARA prompts for simplifications and paraphrasing. This approach aims to enhance comprehension by presenting complex text in a more accessible way, which ultimately helps users to understand the content more efficiently. Finally, SARA seamlessly integrates these results into virtual components within the user's augmented reality environment.
\end{enumerate}


\label{conclusions}
\section{Summary and Future Work}
SARA represents a significant advancement in mixed reality technology for educational support in reading. Using tools such as the HoloLens2 and GPT-4, it provides personalized, real-time assistance tailored to individual reading needs. Through its methodology, including identifying text position, detecting difficulty, and providing targeted support, SARA enhances reading experiences and overcomes comprehension difficulties. Ongoing research and development work aims to improve SARA's capabilities, address limitations, and assess its impact on user engagement and reading comprehension through extensive case studies. SARA demonstrates the potential of mixed reality and advanced language models in redefining reading assistance and advancing inclusive learning environments.


\begin{acks}
We gratefully acknowledge IT-Stiftung Esslingen for their generous support of our hardware lab.
\end{acks}


\bibliographystyle{ACM-Reference-Format}
\bibliography{references}

\end{document}